# Welfare, sustainability, and equity evaluation of the New York City Interborough Express using spatially heterogeneous mode choice models

**Hai Yang**[1] **(**0000-0003-2143-4636**), Hongying Wu**[1]**, Lauren Whang**[2]**, Xiyuan Ren**[1] **(**0000-0002-7719-7695**), Joseph Y. J. Chow**[1*] **(**0000-0002-6471-3419**)**

[1]C2SMARTER Center, Department of Civil & Urban Engineering, New York University Tandon School of Engineering, Brooklyn, NY, USA

[2]Department of Civil & Systems Engineering, Johns Hopkins University, Baltimore, MD, USA

[*]Corresponding author email: joseph.chow@nyu.edu

## Abstract

The Metropolitan Transit Authority (MTA) proposed building a new light rail route called the Interborough Express (IBX) to provide a direct, fast transit linkage between Queens and Brooklyn. An open-access synthetic citywide trip agenda dataset and a block-group-level mode choice model are used to assess the potential impact IBX could bring to New York City (NYC). IBX could save 28.1 minutes to potential riders across the city. For travelers either going to or departing from areas close to IBX, the average time saving is projected to be 29.7 minutes. IBX is projected to have more than 272 thousand daily ridership after its completion (81% higher than reported in the official IBX proposal). Among those riders, more than 58 thousand people (21.4%) would come from low-income households while 185 thousand people (68.2%) would start or end along the IBX corridor. The addition of IBX would attract more than 40 thousand additional daily trips to transit mode, among which more than 16 thousand would be switched from using private vehicles, reducing potential greenhouse gas (GHG) emissions by 30.63 metric tons per day. IBX can also bring significant consumer surplus benefits to the communities, which are estimated to be $0.89 USD per trip. However, the service does not appear to significantly reduce the proportion of travelers whose consumer surpluses fall below 10% of the population average (already quite low).

**Ethical Approval and Conflicts of Interest Statement**: The authors do not have any conflicts of interest to declare.

# 1. Introduction

Public transit systems are a crucial part of urban transportation networks. According to Statista (2023a), an estimated 4.3 billion people are projected to use public transit systems around the world in 2023. Approximately 56% of the world's population has access to public transit, with Asia and Europe providing access to more than 60% of their populations. However, public transit is significantly underdeveloped in the US, with only 35.7% of its population having access. Among commuters, the share of public transit is even lower, with only 16% of US commuters choosing to use public transit frequently (Statista, 2023b). Despite being the largest city and having the largest transit network in the U.S. (APTA, 2022), New York City (NYC) struggles to attract more transit users. The mode share of transit in NYC was 24% in 2019 (NYCDOT, 2020), and it took a significant toll during the pandemic, with ridership only recovering to around 60% of the pre-pandemic level in 2022 (MTA, 2023b). With such low ridership, coupled with fare evasion and aging infrastructure, the Metropolitan Transit Agency (MTA) constantly faces a high budget deficit that impedes its efforts to provide high-quality transit services (MTA, 2023a).

The condition of transit accessibility is less desirable in Brooklyn and Queens, with only one existing subway line directly connecting the two populous boroughs. Figure 1 shows the Transit Connectivity Index (TCI) in Manhattan, Brooklyn, and Queens, respectively. The TCI developed by CNT (2023) measures the transit service levels by using available transit lines and service frequencies in each census tract. By comparing Figure 1a, 1b, and 1c, it is obvious that Brooklyn and Queens have several underserved areas where insufficient public transit services are provided. To alleviate such accessibility issues, Gov. Hochul announced the Interborough Express project in 2022 (MTA, 2024). The proposed IBX will be constructed using an existing freight railway line, and it will operate as a light rail service. According to the preliminary report (MTA, 2023c), over 900,000 residents are located within a 0.5-mile radius along the line, where a high percentage of households belong to disadvantaged groups. As shown in Figure 2, the IBX not only covers many areas with insufficient transit services, but also creates shortcuts across NYC that could significantly reduce existing transit travel times for many trips.

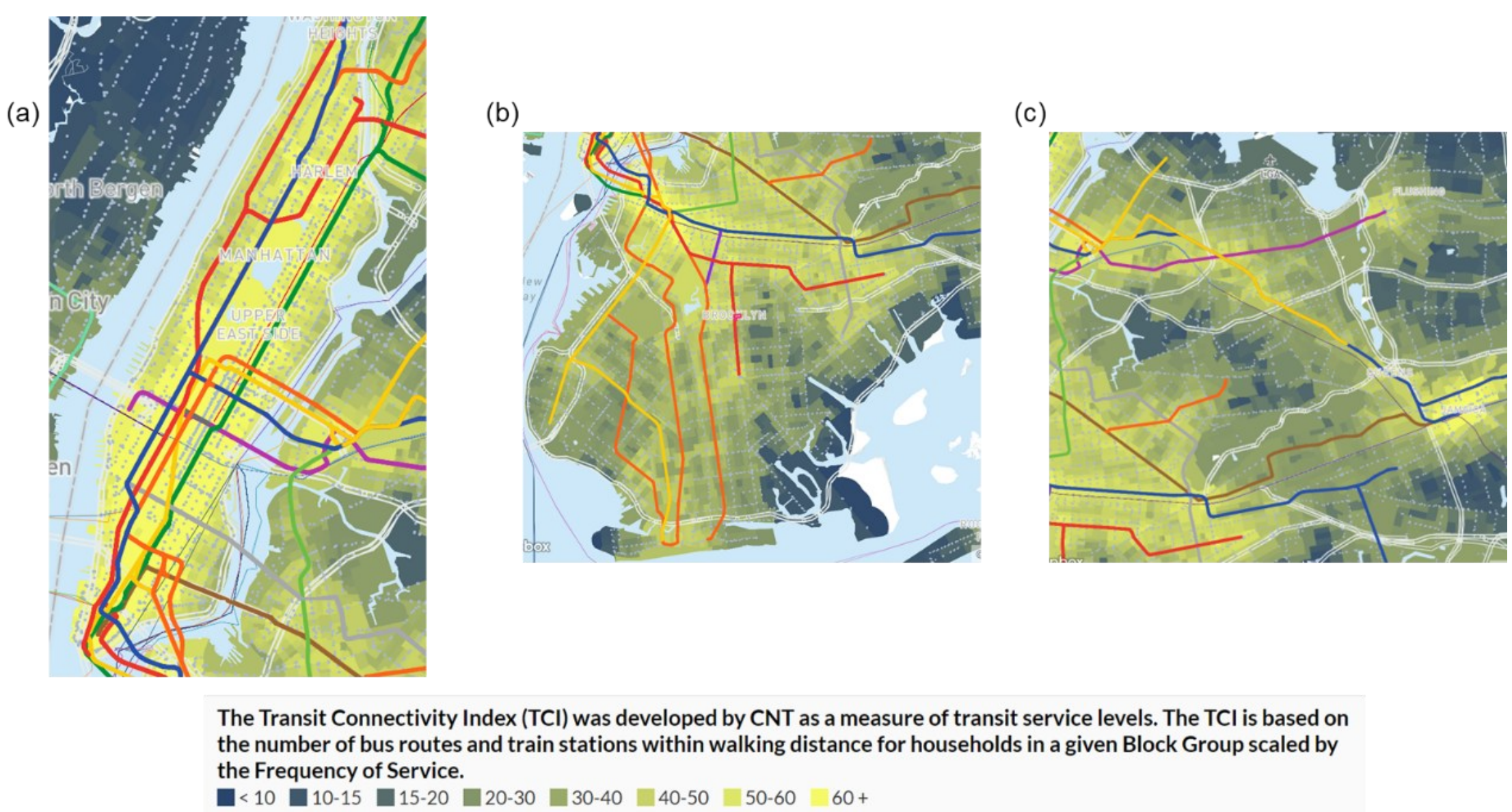

Figure 1. The Transit Connectivity Index (TCI) in (a) Manhattan, (b) Brooklyn, and (c) Queens (source: CNT, 2023)

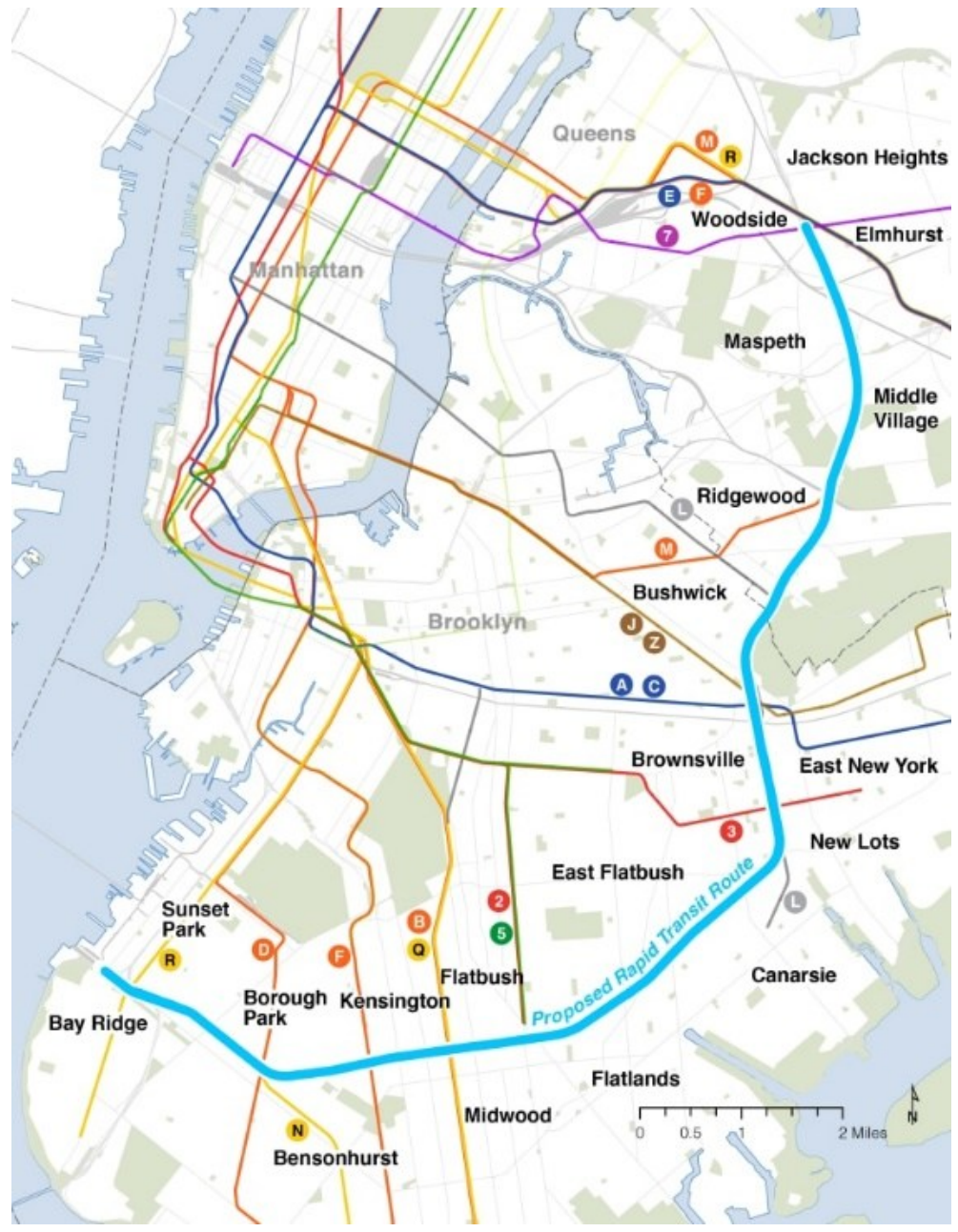

Figure 2. Proposed route design of the IBX

With many potential benefits expected from the IBX, there are limited studies quantifying these benefits comprehensively. There is limited survey data collected specifically for this route, so a regional planning model is needed. However, existing models generally do not capture spatial heterogeneity in taste variation. In other words, the value of time for one income level population segment may be the same regardless of origin and destination or length of trip. However, travelers with trips ending at JFK International Airport will generally have higher values of time than travelers commuting a mile distance somewhere in the outer boroughs. Therefore, an analysis of IBX ridership substitution of existing transit using conventional models may not adequately capture the spatial substitution patterns.

The IBX is expected to deliver substantial benefits, but quantifying these benefits requires a comprehensive understanding of its impacts on traveler behavior across the NYC. Such analysis demands data and models capable of capturing heterogeneity in travelers' sensitivity to travel time, monetary cost, and mode preferences. This heterogeneity arises not only from sociodemographic attributes such as age and income, but also from spatial and built environment factors—including transit accessibility, road network density, land use mix, and employment density—that interact with individual characteristics to shape travel preferences. Existing mode choice models generally do not capture these spatial interactions in taste variation; for example, the value of time for a given income group is often assumed constant regardless of trip origins and destinations. Consequently, these models may misattribute the IBX's ridership and welfare impacts by failing to account for how travelers in different spatial contexts respond to the new service.

We propose to analyze the mode shift from existing options to the IBX using spatially heterogeneous mode choice models estimated from synthetic population data. Unlike traditional mode choice models that allow mainly capture variations across population, spatially heterogeneous models allow taste parameters to vary by both population segment and origin-destination (OD) pair. This flexibility enables the model to capture unobserved spatial factors and complex interactions that traditional global models often overlook. In this study, we employ models with unique coefficients estimated for each OD pair at the U.S. census block group level, further segmented into four mutually exclusive population groups: low-income, non-low-income, seniors, and students. These models were estimated in Ren et al. (2025a), and the parameters are publicly available on Zenodo (Ren & Chow, 2023). The resulting ridership estimates are drawn from a heterogeneous set of travelers and provide a more comprehensive picture of spatial impacts on welfare, equity, and sustainability than conventional models.

The following steps are taken. We present the average transit time savings across different population groups in NYC. Ridership information is generated based on the transit time savings using the mode choice models. In addition, the estimated mode shift from private vehicles to public transit provides insights into the potential benefits of reducing car usage and subsequent savings on vehicle miles traveled (VMT) and approximate greenhouse gas (GHG) emissions. The mode choice model also facilitates the analysis of potential consumer surplus benefits to the entire NYC population. Based on the estimated consumer surplus, an equity analysis is also conducted to further evaluate the impact brought by the IBX.

The structure of this paper is as follows: Section 2 provides a literature review. Section 3 presents the key datasets used in the study. Section 4 presents the involved methodologies; Section 5 presents the analysis results involving transit time savings, ridership impact, consumer surplus benefit, and transit equity analysis; Section 6 concludes this study with discussions.

## 2. Literature review

Transit ridership prediction has long been an important element in urban study. Earlier studies extensively used traditional regression models. Gaudry (1975) utilized a time-series adjusted linear regression model to forecast the transit ridership in Montreal, Canada. Doi & Allen (1986) employed two time series regression models to estimate the transit ridership of a rapid-rail line in the Philadelphia area. Doti & Adibi (1991) applied a regression model with logarithm terms to predict transit ridership in Orange County, California. In addition to regression models, mode choice models have been widely adopted since McFadden's (1974) work. Benjamin et al. (1998) used discrete choice models to forecast paratransit ridership in Winston-Salem, South Carolina. Jovicic and Hansen (2003) incorporated mode choice models in studying public transport ridership in Copenhagen. The classic 4-step model (see McNally, 2007), incorporating mode choice models, remains one of the most prominent ridership forecasting methodologies. Recent advancements in transit ridership prediction leverage novel data-driven and computational methodologies that target the heterogeneity of individual choices. McFadden and Train (2000) incorporated random parameters into the classic discrete choice model to reflect individual's taste. Such taste parameters can be drawn from a general distribution using a conditional approach (Revelt & Train, 2000). However, such models are shown having limited prediction accuracies and having scalability issues (Fox & Gandhi, 2016; Ren & Chow, 2022). Instead of drawing from a parametric distribution, nonparametric logit models directly fit individual-level parameters using the observed data. Swait (2023) proposed an individual parameter logit (IPL) model directly estimating the parameters for individual preferences with a distribution-free approach. However, the IPL model suffers from its computational complexity, preventing its application to larger datasets. Ren & Chow (2022) proposed an agent-based mixed logit (AMXL) model using inverse-optimization (IO) techniques to enable large scale individual-level parameter estimation.

Other types of heterogeneity affecting travel demand have also been studied using different techniques. Eldeeb & Mohamed (2020) used a latent class choice model to reveal how heterogeneity among user groups towards service quality could change travel demand. In Wang et al. (2022), a geographically and temporally weighted regression (GTWR) model was built to reveal the importance of spatial heterogeneity in transit ridership prediction. Orrego-Oñate et al. (2023) proposed a mode choice model with spatial latent classes to capture the effect of spatial features on travel behavior patterns. In recent years, machine-learning algorithms have been widely used in transit ridership predictions. Zhao et al. (2020) provides a comprehensive comparison between well-established discrete choice models and machine learning models in performing travel demand predictions. A common advantage of the machine learning based models over classic discrete choice models is its stronger prediction power, especially in relating spatial heterogeneity to transit demand. Chen et al. (2021) proposed a hybrid model based on random forests (RF) to capture the impact of spatial heterogeneity on transit trip generation. Yap & Cats (2022) used neural network to predict the transit demand during closures based on travelers' origin-destination pairs. Chen et al. (2022) combined a Graph Convolution Network with a stacked Bidirectional Unidirectional Long Short-Term Memory network for enhanced spatial-temporal prediction accuracy. Wu et al. (2024) introduced a Spatial-Temporal Hypergraph Attention Recurrent Network (STHGARN) to capture high-order spatial-temporal relationships, significantly outperforming traditional forecasting methods. Recent studies further incorporate equity aspects into the public transit analysis. Alogdianakis & Dimitriou (2024) proposed a multi objective framework for allocating transit budgets by partitioning urban areas into spatial polygons to balance operational efficiency with social equity. Carleton & Porter (2025) developed a predictive modeling approach for transit travel times to evaluate how infrastructure investments specifically impact transportation equity for underserved populations.

With numerous studies showing the importance of linking spatial heterogeneity with travel demand, it is thus essential to incorporate it in the model. Though machine learning models can capture such linkage, they are generally less interpretable than discrete choice models, which hampers their further applications in welfare related studies. Ren et al. (2025a) proposed a group level agent-based mixed (GLAM) logit model that combines machine learning techniques with discrete choice models. It is based on the IO algorithm used in Ren & Chow (2022). Instead of directly using individual-level observations, individual choices are aggregated into group-level data for parameter fitting and individuals inside the same group are assumed to share the same preferences. In addition, parameter clusters are identified to decompose the fitting process with smaller IO problems. The general logic of the GLAM logit model along with the mixed logit (MXL) and IPL models are shown in Figure 3. The GLAM logit model provides high levels of prediction power while preserving good interpretability. The model allows a more generic approach when doing demand predictions. For example, travelers in the same population group sharing the same origin and destination are likely to have similar preferences. Therefore, spatial heterogeneity can be well captured among future travelers while avoiding the unpredictability of individual behavior. Because the model still follows a discrete choice model structure, it can be easily used to conduct welfare related analysis.

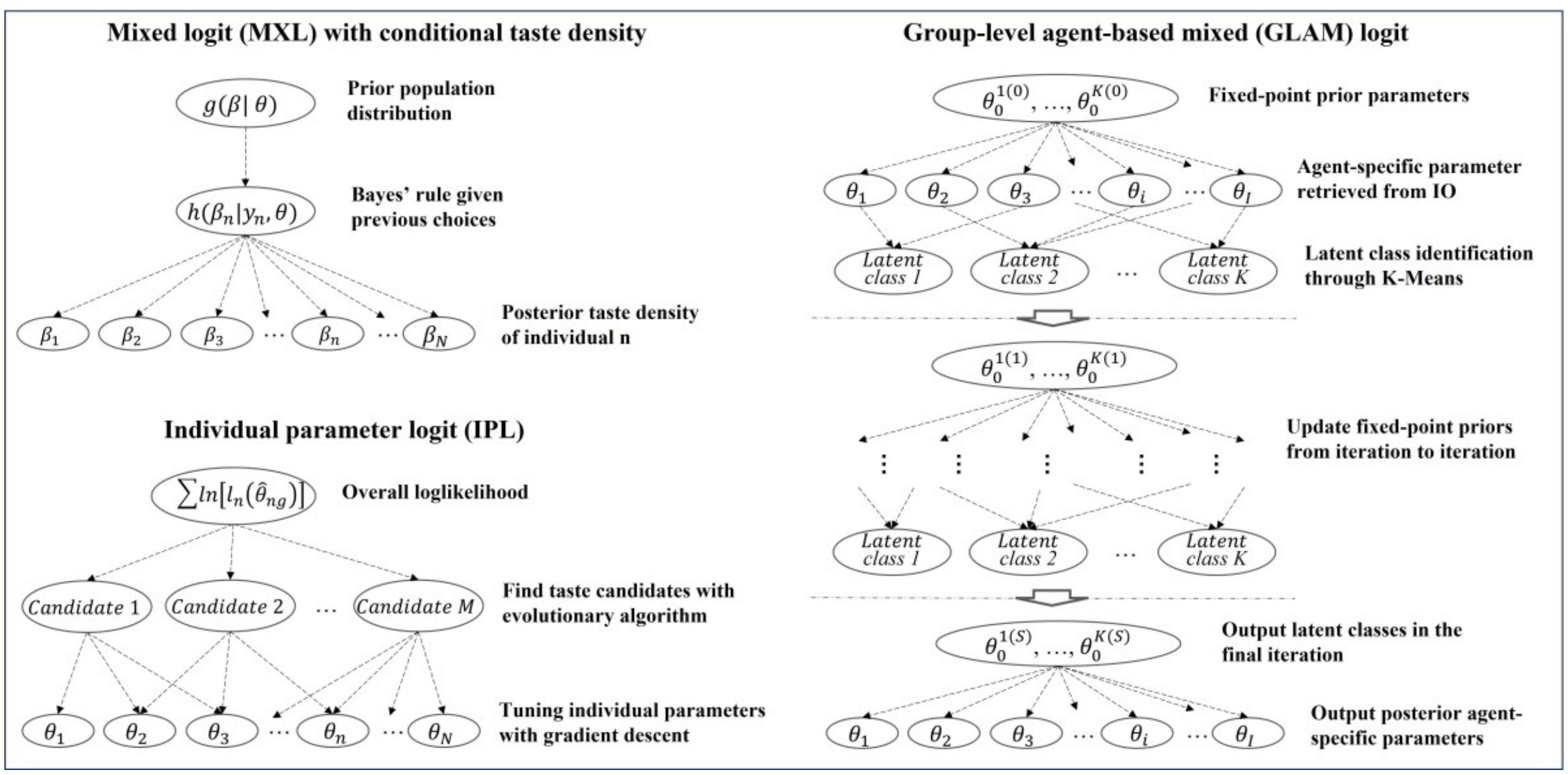

Figure 3. General logics of the MXL, IPL, and GLAM logit models (source: Ren et al., 2025a).

In this study, we investigate the impact of IBX by using the GLAM logit model to capture the spatial heterogeneity among travelers. Not only can the GLAM logit model be used to predict future demand, but the result can also be directly applied to calculate the environmental and social equity impacts, providing a more comprehensive picture of its impact.

## 3. Dataset

The key datasets involved in the projects are the New York City synthetic trip agenda and the datasets generated from a discrete choice model. This section briefly introduces the main concepts and generation process of the two datasets.

### 3.1. NYC synthetic trip agenda

Public data, such as census records, is often spatially aggregated to protect privacy. However, this aggregation prevents the estimation of the granular, agent-level interactions required for this study. Specifically, applying GLAM model coefficients requires detailed trip information, which includes trip-maker attributes and block-level origins and destinations. Such levels of detail are not readily available from public domain. To address this, we use a synthetic dataset that provides the necessary trip-level granularity. The synthetic data is sourced from Replica, a transportation data company. Ren et al. (2025b) conducted a thorough validation comparing the synthetic data with publicly available surveys. Details are attached in Appendix A. The NYC synthetic trip agenda is a dataset that includes a comprehensive list of trip itineraries, illustrating the daily activities and movements of all individuals within NYC on a typical weekday. This subsection summarizes the data generation process, with detailed descriptions available in Wang et al. (2021).

Initially, citywide synthetic household data is created using the American Community Survey (ACS) Public Use Microdata Sample (PUMS). Popgen (MARG, 2016), a population synthesis tool utilizing iterative proportional fitting (IPF) methods (Bar-Gera et al., 2009; Ye et al., 2009; Konduri et al., 2016), extrapolates the PUMS dataset to represent the full-scale NYC household population. The synthetic population includes socio-demographic features such as age, income level, education status, employment industry, home location, and car ownership status.

Work/school locations are then assigned based on the socio-demographic characteristics of the synthesized individuals. By using the Census Transportation Planning Product (CTPP) tabulation from ACS, a home-to-work matrix is constructed to represent the number of residents commuting from home to work/school at the census tract level. These flows, presented as marginal distributions, and the conditional distribution of the work census tract is calculated based on the employed person's industry and home census tract. Work census tracts are assigned reflecting individuals' socio-demographic attributes and the conditional distribution. Figure 4 demonstrates the similarity between the assigned home-work pairs and information from CTPP, showing a high level of concordance.

School location assignments follow a similar logic with some simplifications due to limited information on home-school trip flows. Rather than assigning school based on computed distributions, the closest school is assigned to a student's home census tract based on their education status.

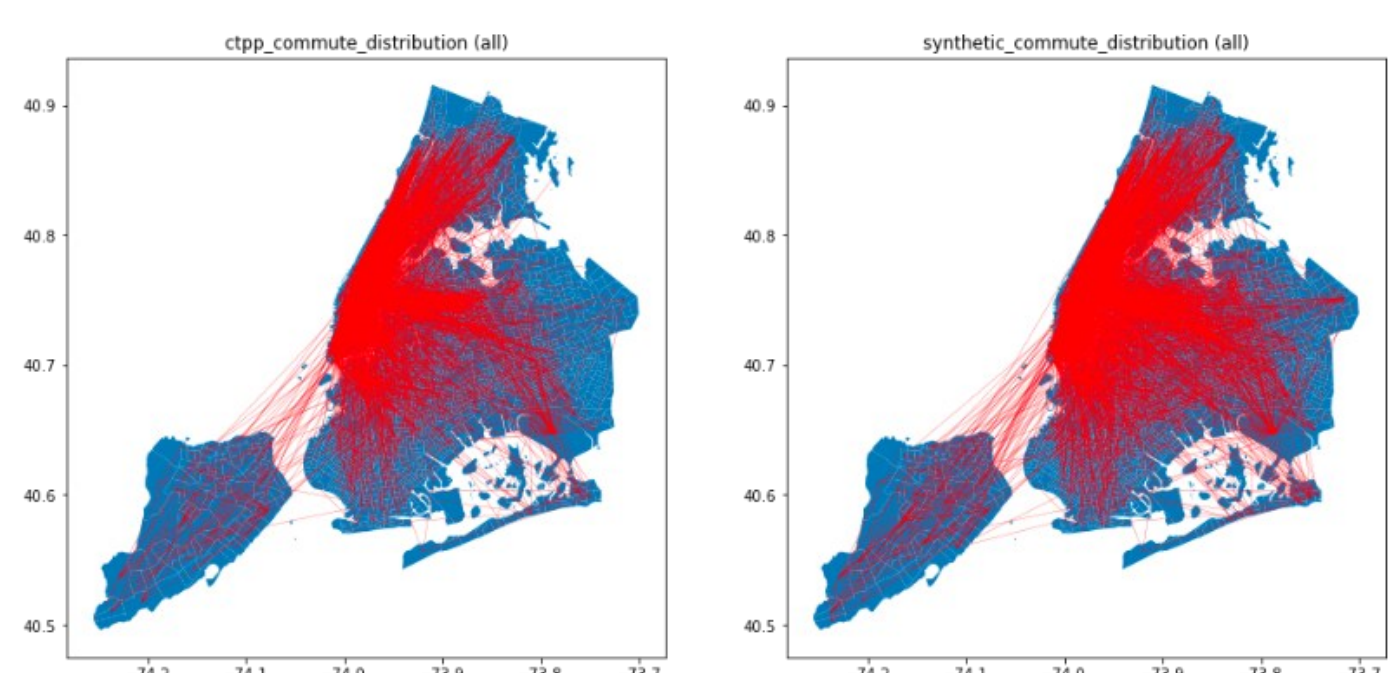

Figure 4. Home-work flow comparison between CTPP (left) and synthetic commute trips (right).

The assigned work/school locations serve as the primary locations in an individual's daily agenda. The MTA mobility survey (MTA, 2019) serves as a reference for generating each person's trip agenda by matching socio-demographic features and primary locations. If no exact match is found in the survey, agendas from the closest matching records are used.

After establishing each person's trip agenda, an appropriate transportation mode is assigned to each trip leg. The block-group level predicted mode share for NYC, provided by Ren et al. (2025a), facilitates this task. This dataset, which includes block-group level mode share for different population segments, is aggregated to the census tract level. Modes are probabilistically assigned to each trip leg based on the mode share. For individuals using private vehicles, a consistency check ensures that both the trip to and from work/school use the same mode.

In conclusion, over 28 million trips are generated, depicting the daily itineraries of over 8.4 million people in NYC.

### 3.2. Datasets from the GLAM logit model

The GLAM logit model is a novel discrete choice model that addresses the heterogeneity of trips on both social-demographic and spatial aspects. For detailed model structure and fitting algorithm, readers can refer to Ren et al. (2025a). The model has heterogenous parameters based on both the population segmentations and the trip origin-destination pairs. According to Ren et al. (2025a), the model performs the best in predicting the outcome of new scenarios (e.g., new transportation services) assuming the same trip profiles in the area, which fits well with the purpose of this study. The model shows a high level of accuracy in predicting mode share percentages for each population group when applied to NYC population. The in-sample accuracy of the GLAM logit model is 96.97%, which is significantly higher than the benchmark models such as the MXL and IPL models. A high level of out-of-sample accuracy is also achieved, reaching above 89%. Again, the GLAM logit model outperforms the benchmark models by large margins.

The GLAM logit model is designed for ubiquitous datasets. Therefore, the model provides better results when fitted with larger datasets. A parameter set of the GLAM logit model is fitted using Replica's New York State (NYS) population dataset, which contains more than 53.5 million trips. Six transportation modes are involved: private vehicle, public transit, on-demand mobility service, biking, walking, and carpool. The trips are grouped by the population segmentations and the census block level origin-destination (OD) pairs. For the population segmentations, a hierarchical categorization is applied to ensure mutual exclusivity. Individuals aged 65 and older are first identified as seniors, followed by individuals enrolled in school who are classified as students. The remaining population is subsequently divided into low-income and not–low-income groups according to the Federal Poverty Guidelines. As a result, 120,740 trip groups are identified, with each group having a unique parameter set, resulting in over 2.5 million parameters. The parameters included in the GLAM logit model are summarized in Table 1.

Table 1. Parameters of the GLAM logit model per trip group

| ***Transportation mode*** | **Trip elements** |
|---|---|
| *Private vehicle* | $\theta_{auto_{tt},i}, \theta_{cost,i}, asc_{driving,i}$ |
| *Public transit* | $\theta_{transit_{at},i}, \theta_{transit_{et},i}, \theta_{transit_{ivt},i}, \theta_{transit_{nt},i}, \theta_{cost,i}, asc_{transit,i}$ |
| *On-demand mobility service* | $\theta_{auto_{tt,i}}, \theta_{cost,i}, asc_{on_demand,i}$ |
| *Biking* | $\theta_{nonvehicle_{tt},i,}, \theta_{cost,i}, asc_{biking,i}$ |
| *Walking* | $\theta_{nonvehicle_{tt},i,}, \theta_{cost,i}, asc_{walking,i}$ |
| *Carpool* | $\theta_{auto_{tt},i,}, \theta_{cost,i}, asc_{carpool,i}$ |

$\theta_{auto_{tt},i}$ measures the disutility of car travel time for each trip group $i$. In this model, driving private vehicles, using on-demand mobility service, and using carpool share the same travel time parameter. $\theta_{cost,i}$ measures the monetary aspect of the trip, and all modes share the same parameter for each trip group $i$. $\theta_{transit_{at},i}$, $\theta_{transit_{et},i}$, and $\theta_{transit_{ivt},i}$ measure the disutility of transit access time, transit egress time, and transit in-vehicle time for trip group $i$. $\theta_{transit_{nt}}, i$ measures the negative impact caused by the increased number of transfers. $\theta_{nonvehicle_{tt},i}$ measures the disutility of trip time that does not involve any automobile. In addition to travel time and cost related parameters, each mode has its constant value labelled as $asc$ that represents the general mode preference among trip makers for trip group $i$.

The value of time (VOT) of trip makers can be estimated by dividing time related parameters by cost related parameters (e.g., $\theta_{auto_{tt},i}/\theta_{cost_i}$). Table 2 summarizes the average VOT among population groups in NYS and NYC. Significantly higher VOTs for the NYC population generated from the GLAM logit model align with the higher income level in NYC. When looking into the geographic distribution of VOT, distinctive patterns can be observed as shown in Figure 5. Higher levels of VOT are clustered in urban areas, especially in Manhattan. Such patterns match the empirical knowledge of the income level distributions.

Table 2. Average value of time (VOT) of population segments. (source: Ren et al., 2025a)

| ***Population segment*** | **Average VOT in NYS** | **Average VOT in NYC** |
|---|---|---|
| *Notlowincome* | \$18.11/hour | \$28.53/hour |
| *lowincome* | \$8.92/hour | \$9.62/hour |
| *senior* | \$12.07/hour | \$13.23/hour |
| *student* | \$9.94/hour | \$11.65/hour |

The two datasets form the fundaments of the transit analysis process. In the following section, we present the methodology this analysis adapts and how these datasets are used.

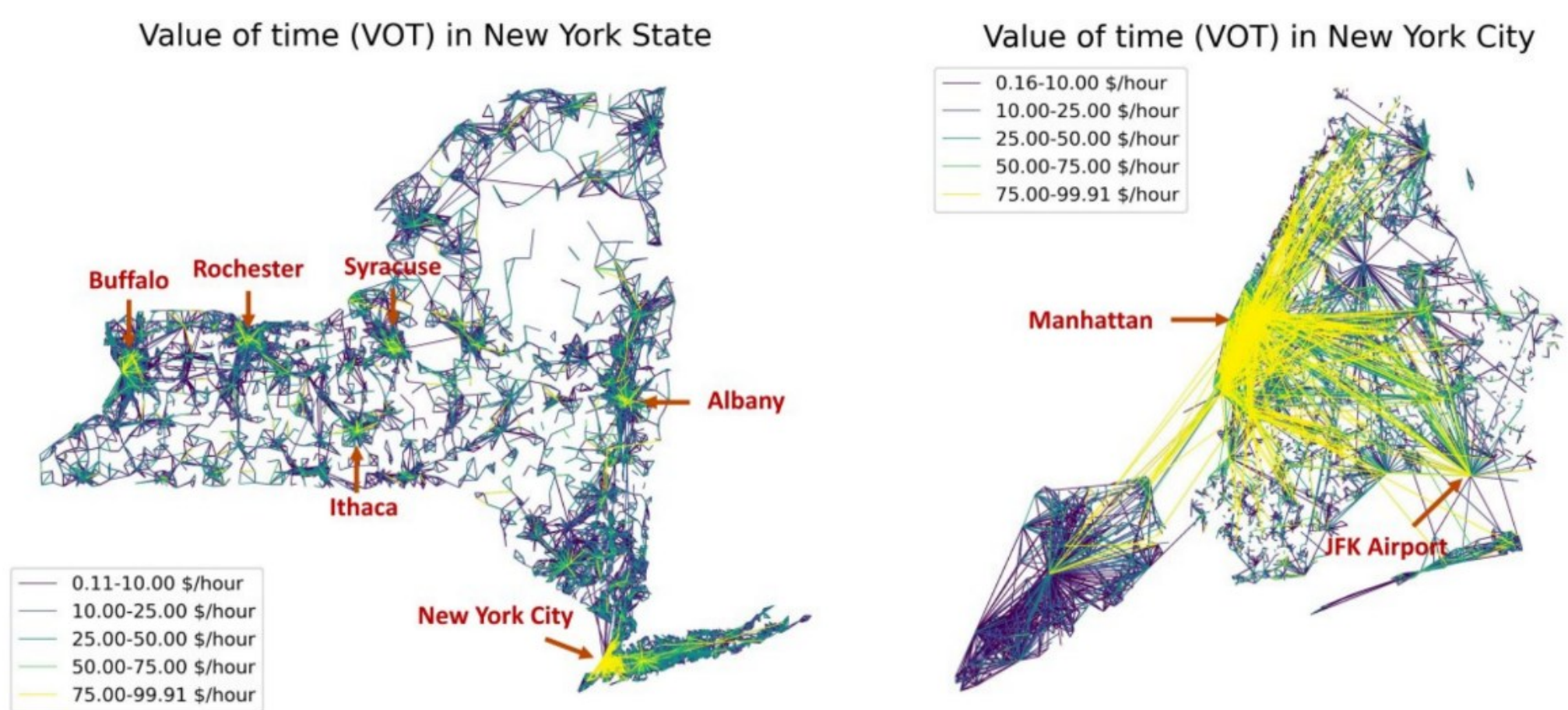

Figure 5. Geographic distribution of VOT in NYS (left) and NYC (right). (source: Ren et al., 2025a)

# 4. Methodology

The study comprises four principal steps to conduct the required analysis of the proposed IBX: synthetic workday service schedule creation, transit time saving matrix generation, ridership and consumer surplus calculation, and equity analysis.

The analytical framework follows a sequential logic to translate IBX implementation into social outcomes, as illustrated in the Figure 6 flowchart. The process begins with the GTFS generator, which integrates information from the IBX proposal with existing subway service. This establishes the foundational service addition required for the study.

Next, the SBB transit time module compares this synthetic schedule against current MTA GTFS data to generate a transit time saving matrix. This step quantifies the direct operational impact of the IBX on travel times across the network. These time savings, along with the NYC trip agenda and fitted model parameters, serve as primary inputs for the ridership and consumer surplus calculation. This process identifies mode shift behavior and estimates how many travelers transition to the IBX based on improved utility.

Finally, the analysis yields three primary outputs: IBX ridership, mode shift, and consumer surplus. The individual welfare gains, measured as consumer surplus, are then processed through the equity analysis stage. This step determines the equity impact by assessing how these benefits are distributed across different socio-demographic groups, specifically focusing on underserved populations.

### 4.1 Synthetic service schedule

The latest proposal (MTA, 2024) suggests constructing the IBX along an existing freight rail corridor that stretches 14 miles with 17 potential stops, offering a direct connection between Queens and Brooklyn in under 39 minutes as a light rail service. During peak hours, the light rail is expected to operate with a 5-minute headway. Details such as full day schedules are not yet provided.

We develop our version of the full workday IBX schedule by analyzing the temporal distribution of trips from the trip agenda to determine operation periods. The number of trips by departure hour is shown in Figure 7. Table 3 summarizes the five schedule intervals identified with corresponding headways, where morning peak and evening peak headways are directly obtained from the proposal, and the others are determined based on the general operation status of existing subway lines (e.g., the 7 line). This schedule represents a reasonable design to serve as a basis for comparison against a benchmark without the IBX.

With these schedule intervals and headways, a full day schedule is created. To estimate travel time between stops, we use the average travel speed calculated from the total length of the IBX line divided by the proposed total travel time. Travel times are assumed identical in both directions, assuming identical operating conditions. A GTFS builder from the National Rural Transit Assistance Program (RTAP, 2024) is used to construct the schedule in spreadsheets and convert it to GTFS files of the IBX, which are available in the Zenodo repository (Yang, 2023).

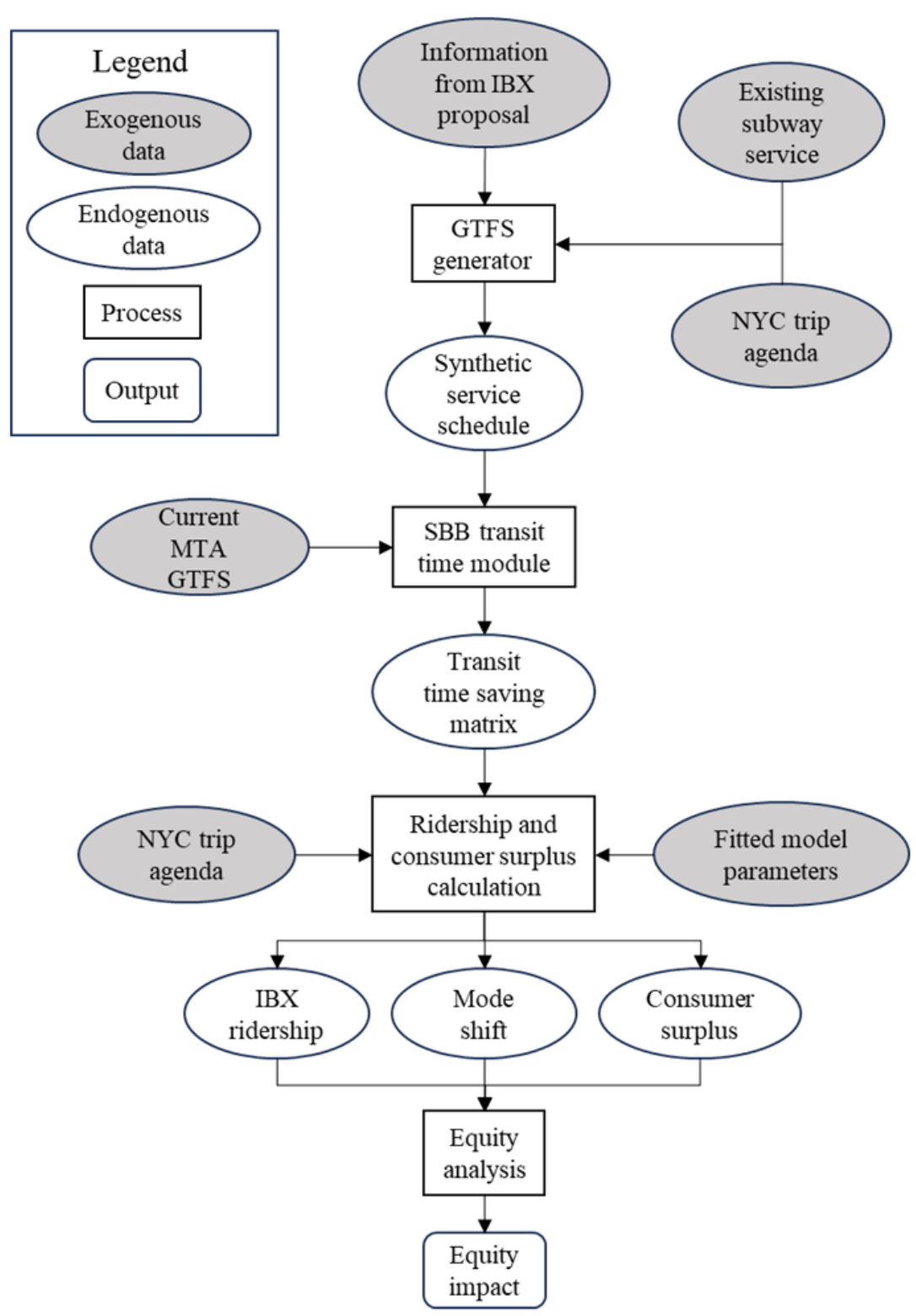

Figure 6. Flowchart of the study.

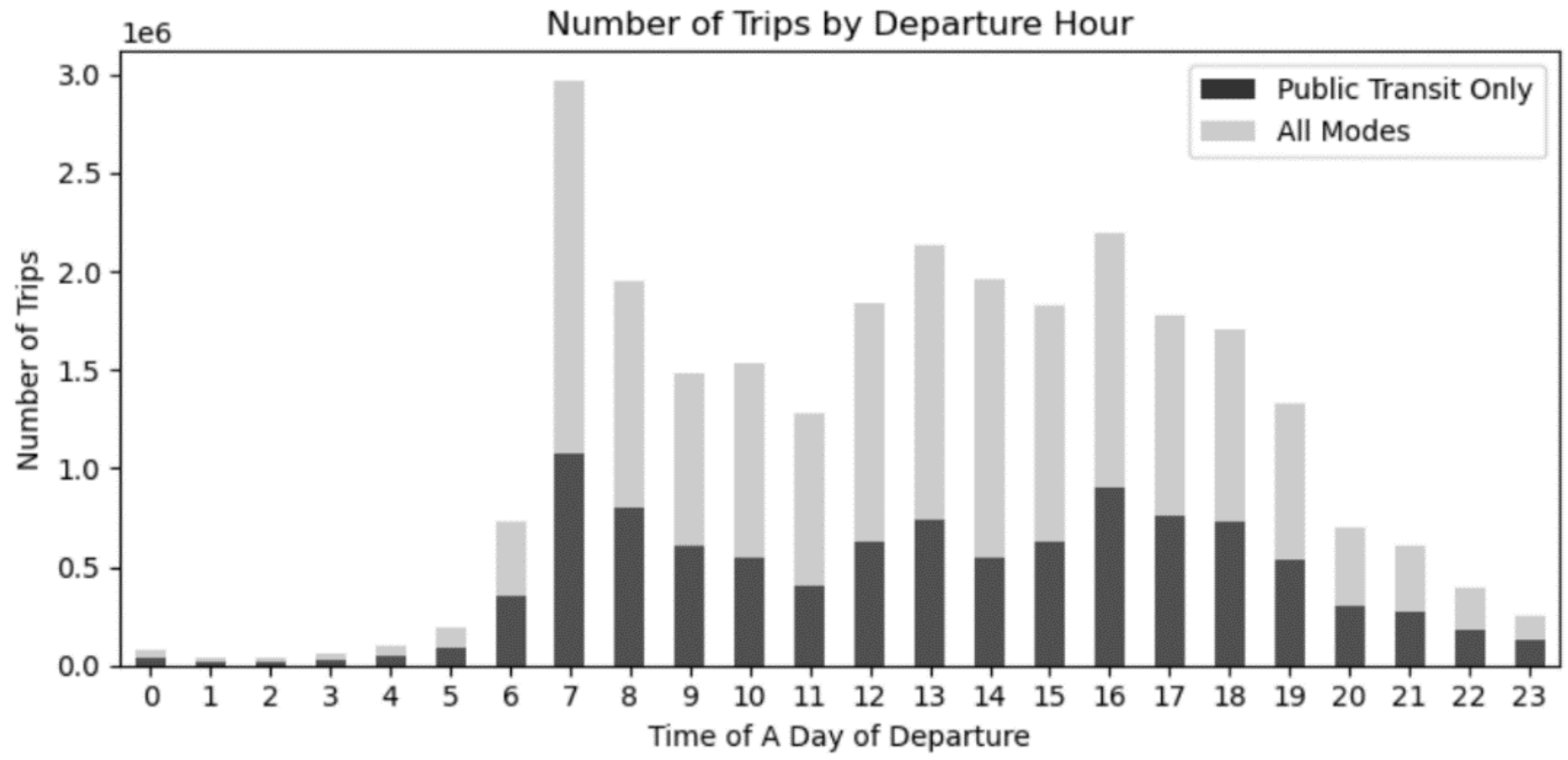

Figure 7. Number of trips by departure hour.

Table 3. Schedule intervals and corresponding headways.

| ***Schedule interval*** | **Time span** | **Headway** |
|---|---|---|
| *Morning peak* | 6 AM – 9 AM | 5 minutes |
| *Mid-day* | 9 AM – 4 PM | 10 minutes |
| *Evening peak* | 4 PM – 8 PM | 5 minutes |
| *Evening* | 8 PM – 11 PM | 10 minutes |
| *Early morning* | 11 AM – 6 AM | 20 minutes |

### a. Transit time matrix

After the IBX schedule and corresponding GTFS files are generated, potential travel time savings OD pairs using transit can be estimated. The 2010 census tract serves as the basic geographic unit for transit time calculation. The centroid of each census tract represents all locations within the area, and the centroid-to-centroid transit travel time represents the average transit travel time for all trips occurring between the two areas.

The Skim Matrices extension by Swiss Federal Railways (SBB, 2024) is employed to estimate transit travel time for all census tract OD pairs. This extension generates trip samples with provided OD locations and start times, simulated in MATSim, with the expected realized travel time as the final output. The extension is available at SBB (2024). For each OD pair, the access time, egress time, and in-vehicle travel time of the transit trip are documented.

Initially, the extension runs in the MATSim environment using the MTA GTFS file from March 2023 to produce the current transit travel time matrix for all census tract OD pairs in NYC. The IBX GTFS files are then integrated with the original MTA GTFS files to update the transit system for the MATSim simulation environment. Running the extension again in this updated environment yields the "new" transit travel time matrix. The process is repeated under five different scenarios for each simulation environment, resulting in a total of 10 transit time matrices, each incorporating all three travel time components.

### b. Ridership, mode share shift, and consumer surplus change calculation

We assume that attributes associated with transit travel time are the only variables affected by the implementation of the IBX service, which include access time, egress time, and in-vehicle time. Holding other factors constant (such as socio-demographic characteristics, travel costs, and travel times for other modes), the changes in ridership attributable to transit travel time savings are evaluated using the elasticity measure between travel time and transit market share.

We define the benchmark travel time between census tract $i$ and $j$ prior to IBX implementation as $t_{ij}$, the change of travel time as $\Delta t_{ij}$, the benchmark mode share of transit as $P_{ij}$, and change of mode share of transit as $\Delta P_{ij}$. The point elasticity $e$ between mode share of transit and transit travel time can be written as Eq. (1).

$$e_{ij} = \frac{\Delta p_{ij}/p_{ij}}{\Delta t_{ij}/t_{ij}} = \frac{\Delta p_{ij}}{\Delta t_{ij}} \times \frac{t_{ij}}{p_{ij}} \tag{1}$$

When we are to predict the percentage change of $p_{ij}$ induced by a percentage change of $t_{ij}$, we can use Eq. (2).

$$\frac{\Delta p_{ij}}{p_{ij}} = e_{ij} \frac{\Delta t_{ij}}{t_{ij}} \tag{2}$$

If the travel time elasticity of the market share of transit is $-2$, a 10% increase in travel time causes the share to fall by 20%. Therefore, if the original share of transit is 20%, the new transit share caused by the lengthened transit trip would become 16%.

For the classic multinomial logit (MNL) model, Eq. (1) results in Eq. (3).

$$e_{ij} = \frac{\Delta p_{ij}}{\Delta t_{ij}} \times \frac{t_{ij}}{p_{ij}} = \theta\left(1 - p_{ij}\right)t_{ij} \tag{3}$$

where $\theta$ is the parameter associated with travel time. Gensch & Recker (1979) provides the detailed explanation of the derivation. By rewriting Eq. (3), the change of mode probability is therefore shown in Eq. (4).

$$\frac{\Delta p_{ij}}{p_{ij}} = \theta\left(1 - p_{ij}\right)\Delta t_{ij} \tag{4}$$

In this study, $\Delta t_{ij}$ consists of access time $\Delta t_{at,ij}$, egress time $\Delta t_{et,ij}$, and in-vehicle time $\Delta t_{ivt,ij}$. Therefore, Eq. (4) can be rewritten to Eq. (5).

$$\frac{\Delta p_{ij}}{p_{ij}} = \left(1 - p_{ij}\right)(\theta_{at,ij}\Delta t_{at,ij} + \theta_{et,ij}\Delta t_{et,ij} + \theta_{ivt,ij}\Delta t_{ivt,ij}) \tag{5}$$

where $\theta_{at,ij}$, $\theta_{et,ij}$, and $\theta_{ivt,ij}$ correspond to the parameters for access time, egress time, and in-vehicle time.

Since the GLAM model uses a similar formation of the MNL model for each trip group, Eq. (5) can be directly adapted to calculate the mode share change. The parameter set presented in section 3.2 is based on trip groups defined by census block level OD pairs. Since we use census tract level OD pairs for our analysis, we aggregate the parameters into census tract level by calculating the trip number weighted average. The three travel time associated parameters involved in the model are: $\theta_{transit_{at},i}$, $\theta_{transit_{et},i}$, and $\theta_{transit_{ivt},i}$ for trip group $i$. The provided parameter set is based on trip groups defined by census block level OD pairs. Since we use census tract level OD pairs for our analysis, we aggregate the parameters into census tract level by calculating the demand-weighted average. Using the same notations introduced in section 3.2, the transit mode share change is calculated using Eq. (6).

$$\begin{aligned}\frac{\Delta p_{transit,i}}{p_{transit,i}} = \left(1 - p_{transit,i}\right)(&\theta_{transit_{at},i}\Delta t_{transit_{at},i} + \theta_{transit_{et},i}\Delta t_{transit_{et},i} \\ &+ \theta_{transit_{ivt},i}\Delta t_{transit_{ivt},i})\end{aligned} \tag{6}$$

where $p_{transit,i}$ and $\Delta p_{transit,i}$ represent the original transit mode share and the change of transit mode share of trip group $i$. $p_{transit,i}$ can be directly calculated from the synthetic agenda. $\Delta t_{transit_{at},i}$, $\Delta t_{transit_{et},i}$, and $\Delta t_{transit_{ivt},i}$ represent the access travel time change, egress time change, and the in-vehicle time change of transit trips in trip group $i$ with the IBX service. The time change elements can be directly calculated using the generated transit time. Since the new transit mode share $p'_{transit,i}$ is simply the old transit mode share plus the transit mode change, $p'_{transit,i}$ is therefore shown in Eq. (7).

$$p'_{transit,i} = p_{transit,i}\left[1 + \left(1 - p_{transit,i}\right)\left(\theta_{transit_{at},i}\Delta t_{transit_{at},i} + \theta_{transit_{et},i}\Delta t_{transit_{et},i} + \theta_{transit_{ivt},i}\Delta t_{transit_{ivt},i}\right)\right] \quad (7)$$

For other modes, the market share would change at the same scale. Therefore, the new mode share of mode $j$ in trip group $i$ can be calculated using Eq. (8).

$$p'_{j,i} = \frac{1 - p'_{transit,i}}{1 - p_{transit,i}} p_{j,i} \quad (8)$$

We assume that the volume of trips within each trip group remains unchanged before and after the introduction of the IBX service. Furthermore, it is assumed that the pattern of trips remains consistent once the IBX becomes accessible to the public. Consequently, the revised transit ridership for each trip group can be directly determined by multiplying the total volume of trips within that group by the corresponding new transit mode share in which there is a transit travel time improvement. If no transit time savings are realized due to the IBX service for a particular trip group, it is assumed that transit ridership for that group remains unchanged, with no members of the group using the transit route that includes the IBX. In such instances, the trip group in question would contribute zero ridership to the IBX. Conversely, if a reduction in transit travel time is observed with the addition of the IBX service for a given trip group, the transit route that includes the IBX becomes more attractive. It is then assumed that all individuals opting for transit will choose routes that incorporate the IBX. Essentially, all transit trips within such a trip group are considered as contributing to IBX ridership. Following this rationale, the aggregate IBX ridership is thus the sum of all new transit trips within groups that benefit from reduced transit times.

Moreover, when a trip group benefits from shorter transit travel times, all associated trips gain from the enhanced convenience of transit, regardless of the final mode of transport selected. This advantage can be quantified by measuring the increase in consumer surplus as a welfare measure resulting from the improved transit service. If using a MNL model, the expected consumer surplus (CS) for each trip in trip group $i$ can be written as Eq. (9).

$$E(CS_i) = \left(\frac{1}{\theta_{cost,i}}\right)\ln\left(\sum^{J} e^{V_j}\right) = \left(\frac{1}{\theta_{cost,i}}\right)\ln\left(\frac{e^{V_{transit,i}}}{p_{transit,i}}\right) \quad (9)$$

where $J$ is the whole set of transportation modes, $V_j$ is the utility of mode $j$, and $\theta_{cost,i}$ is the monetary value per utility. By implementing the formulation of MNL, the expected consumer surplus can be calculated by using only transit related terms. Using Eq. (9), the expected consumer surplus change for each trip in trip group $i$ can be calculated as Eq. (10).

$$\Delta E(CS_i) = \left(\frac{1}{\theta_{cost,i}}\right)\left(\ln\left(\frac{e^{V'_{transit,i}}}{p'_{transit,i}}\right) - \ln\left(\frac{e^{V_{transit,i}}}{p_{transit,i}}\right)\right)$$
$$= \left(\frac{1}{\theta_{cost,i}}\right)\left[\ln\left(\frac{p_{transit,i}}{p'_{transit,i}}\right) + \theta_{tranist_{at},i}\Delta t_{tranist_{at},i} + \theta_{tranist_{et},i}\Delta t_{tranist_{et},i} + \theta_{tranist_{ivt},i}\Delta t_{tranist_{ivt},i}\right] \quad (10)$$

By multiplying the expected consumer surplus change per trip with the number of trips in trip group $i$, the total consumer surplus gain brought by the IBX service can be evaluated, along with the new consumer surplus after operating the IBX.

### c. Equity analysis

The calculated changes in ridership and consumer surplus are further utilized for equity analysis, which assesses whether the introduction of the IBX service leads to increased equity in mobility. For this purpose, consumer surplus is employed to evaluate the equity impact of the IBX. The methodologies for disparity and insufficiency analysis draw on approaches similar to those used by Martens et al. (2022). This subsection outlines the inputs and formulas used in the disparity and insufficiency analysis.

We first assess the level of consumer surplus disparity between the low-income population and the entire population before and after the IBX's introduction. To measure this disparity, we calculate the ratio of the average consumer surplus of low-income trip makers to the average consumer surplus of the entire population. We define this disparity ratio as consumer surplus disparity index (CSDI), which can be calculated as shown in Eq. (11).

$$CSDI = \frac{\sum_{i\in I^l} E(CS_i)d_i}{\sum_{i\in I^l} d_i} / \frac{\sum_{i\in I} E(CS_i)d_i}{\sum_{i\in I} d_i}, \quad (11)$$

where $I^l$ is the set of low-income trip groups, $I$ is the set of all trip groups, and $d_i$ is number of trips in trip group $i$. A ratio above one means that the low-income riders are benefited more from mobility services than the entire population. By calculating the CSDI before and after running the IBX using the computed consumer surplus, we can observe whether the IBX provides higher equity level. Increased CSDI value means positive equity impact.

While disparity can serve as a useful equity indicator, it may not fully capture differences within groups. In the context of the IBX analysis, the averaging of consumer surplus gains for residents living near the stations and those residing farther from the IBX line could obscure significant within-group variations. To count such internal differences, we define the consumer surplus insufficiency index (CSII) as Eq. (12).

$$CSII = \frac{1}{\sum_{i\in I^l} d_i}\sum_{i\in I^l}\left(\max\left(0, \frac{z - E(CS_i)}{z}\right)\right)^2 d_i, \quad (12)$$

where $z$ is the sufficiency threshold. In this study, we vary $z$ to be 10% and 50% of the average consumer surplus of all trip groups for a more thorough equity comparison, which are the values suggested by Martens et al. (2022). The CSII addresses a critical question: to what extent does the consumer surplus of the low-income population fall below a defined sufficiency threshold. A higher CSII value indicates a larger proportion of the trips made by low-income population that does not have an adequate consumer surplus.

## 5. Results

In this sections, three parts of the analysis results are presented: transit travel time saving analysis, ridership and mode switch effect analysis, consumer surplus change and equity analysis.

#### 5.1.1. Transit travel time savings

Transit travel time matrices based on the current transit system and the transit system with IBX services are generated to calculate the transit travel time savings on census tract level OD pairs. Since 5 operation periods are defined for IBX, 5 different time saving matrices are generated. The transit travel time saving we use here is the total saving of access time, in-vehicle time, and egress time. Trip volume weighted averages are used to aggregate the time savings throughout the day. We only involve the trip volumes (from all modes) from OD pairs that would benefit from the introduction of IBX.

Figure 8(a) and 8(b) present the average transit travel time saving by departure and arrival census tract. The area encircled by the grey line is the 0.5-mile buffer zone of the IBX line, which we name the *IBX corridor*. The areas that benefit the most from the IBX in terms of transit travel time savings are the East Flatbush community plus surroundings in Brooklyn and the Maspeth community in Queens. Both areas are located inside the IBX corridor, and they are two of the areas with less transit connectivity as indicated in Figure 1. With the introduction of IBX, higher transit accessibility is provided to these currently underserved communities.

Table 4 presents the average transit travel time savings experienced by different trip groups. Overall, the average transit travel time saving is 28.1 minutes for all the trips benefited from using IBX. When only counting the trips made by low-income population, the average transit travel time savings is 25 minutes. Unsurprisingly, trips either originating from or destined for the census tracts within the IBX corridor would experience higher transit travel time savings, averaging 29.7 minutes.

Table 4. Average transit travel time savings of affected trips.

| Trip groups | Average transit travel time saving per trip |
|---|---|
| Entire population | 28.1 minutes |
| Trip makers from low-income population | 25.0 minutes |
| Departing from/arriving to the IBX corridor | 29.7 minutes |

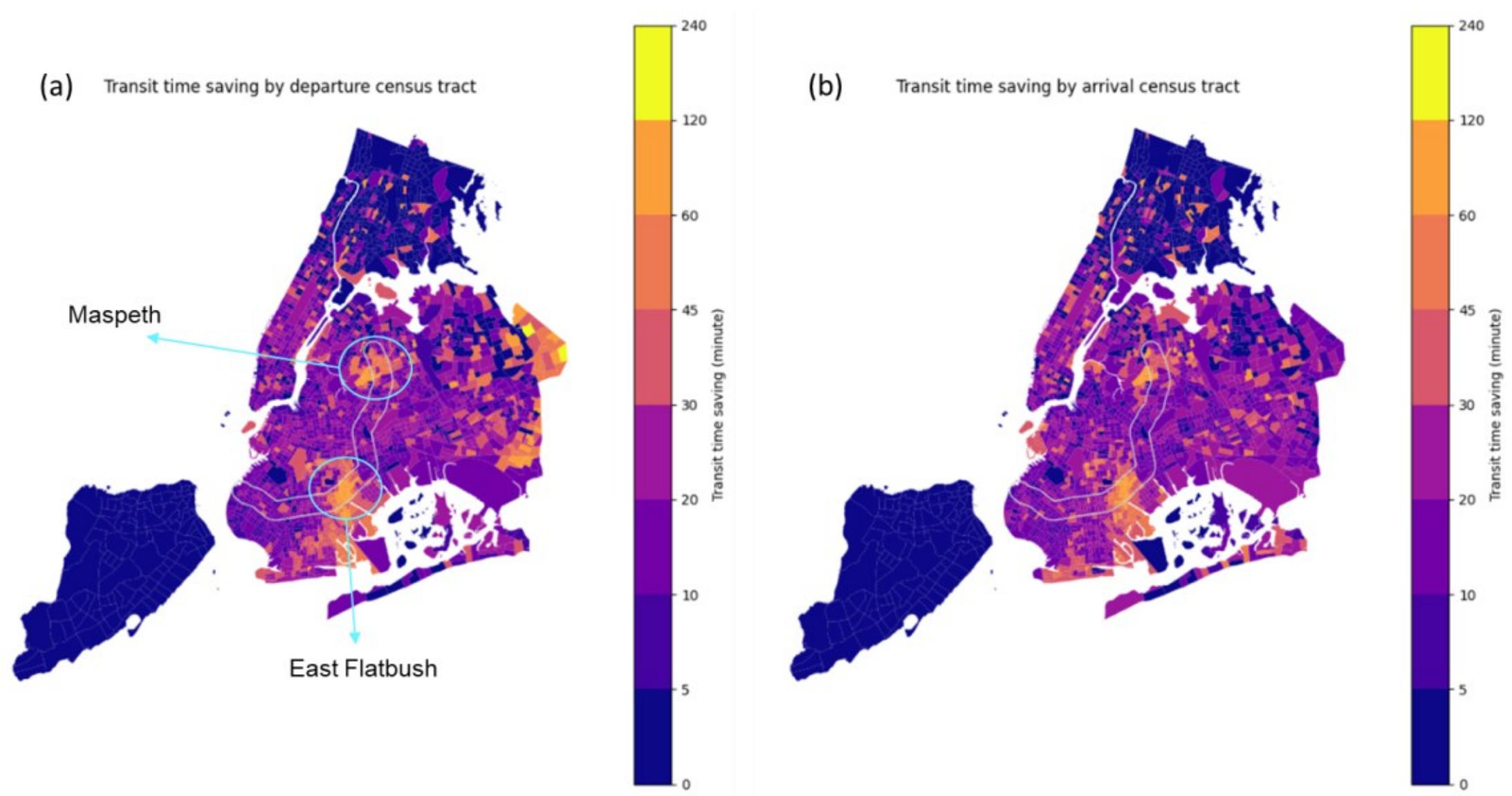

Figure 8. Average transit travel time saving by census tract of (a) departure and (b) arrival; IBX corridor highlighted by grey boundary.

#### 5.1.2. Ridership and mode switch effect analysis

By applying the method described in section 4.3, the IBX daily ridership can be estimated based on the impact of savings on access time, in-vehicle time, and egress time on the mode share of transit. The IBX ridership is estimated by counting all the mode share model transit trips in the trip groups that experience shorter transit travel time. Table 5 summarizes the estimated IBX daily ridership generated from different trip groups. The total IBX ridership is estimated to be more than 272 thousand on a regular weekday. Among them, 58 thousand trips are made by the low-income population segment, which is 21.4% of the daily IBX ridership. Trips that either start from or end within the IBX corridor contribute to 68.17% of the IBX ridership, which is expected because of the added convenience brought by IBX.

Figure 9 and 10 show the IBX ridership by origin and destination census tracts generated by the entire population and only the low-income population. The geographic distribution patterns of the IBX ridership are similar between trips made by the entire population and the low-income population. Aligning with the results shown in Table 5, trips either start from or end at the areas within or near the IBX corridor are big contributors to the IBX ridership. Interestingly, trips that associated with areas relatively far away from the IBX corridor, such as the JFK airport and midtown Manhattan, would also use IBX heavily, indicating its value in providing better transfer services to other parts of the city.

Table 5. Estimated IBX daily ridership.

| Trip groups | Estimated IBX daily ridership (% total) |
|---|---|
| Whole NYC | 272,153 (100%) |
| Trip makers from low-income population | 58,236 (21.40%) |
| Departing from/arriving to the corridor | 185,519 (68.17%) |

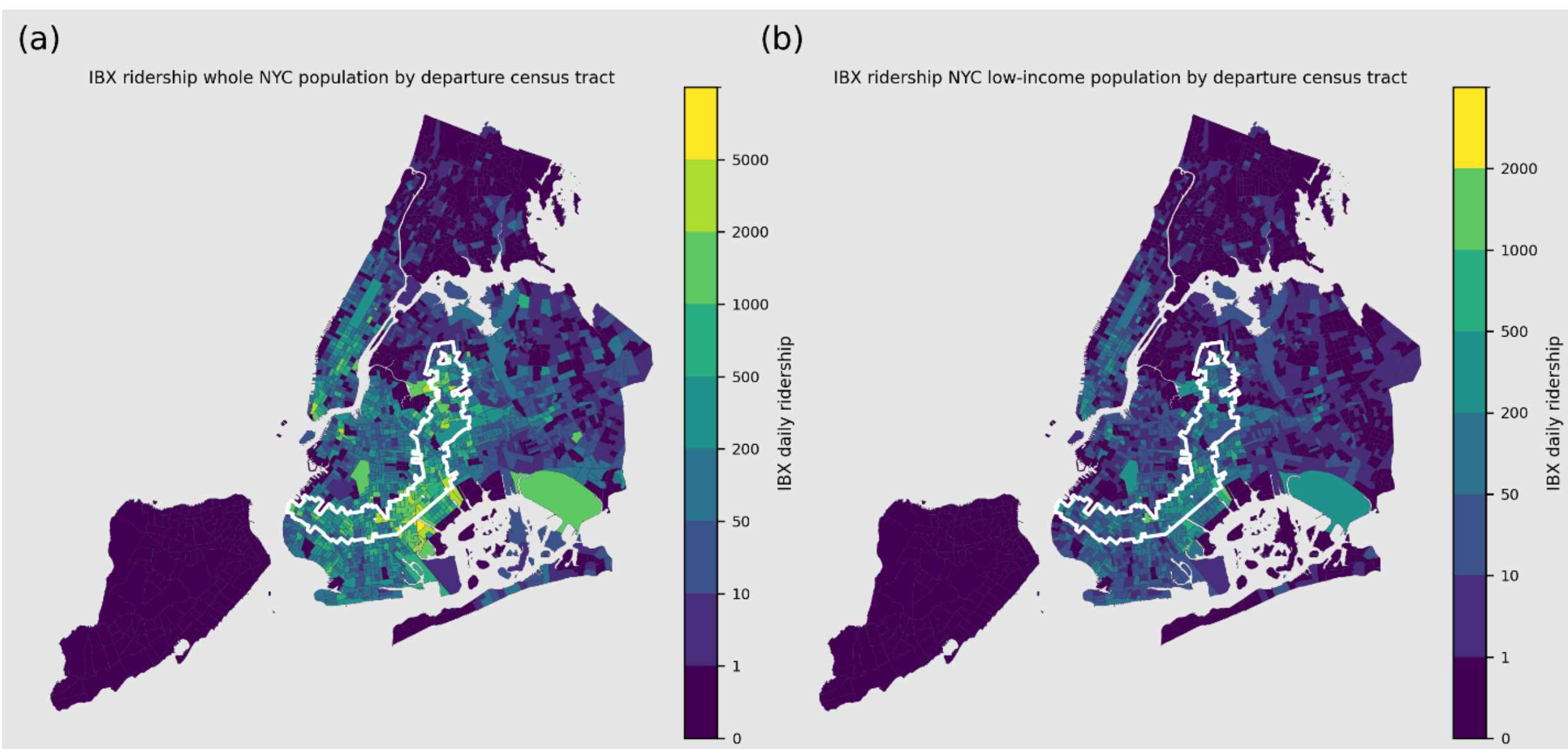

Figure 9. IBX daily ridership by origin census tract generated by (a) entire population and (b) low-income population.

The shortened transit travel time would make transit trips more attractive, potentially alluring more people to shift from other modes to transit. By applying Eq. (7) and (8), the mode shares can be calculated for each trip group and the number of trips switched from one mode to another can be evaluated. Table 6 lists the total transit ridership change and the number of trips switched from private vehicle to transit. 40.2 thousand additional transit trips are generated from IBX, among which 16.4 thousand trips previously used private vehicles. Assuming each vehicle trip require one vehicle on road and all vehicles use internal combustion engine, the potential yearly greenhouse gas emission (GHG) saving could be 30.63 metric tons by applying 400 grams per mile GHG emission standard (EPA, 2023). The estimated emission savings do not account for effects brought by roadway congestion relief, induced demand, or the ongoing electrification of the vehicle fleet. Furthermore, the analysis does not incorporate dynamic speed profiles specific to New

York City. Consequently, these estimations provide a general scale of the emission reduction potential rather than a precise forecast.

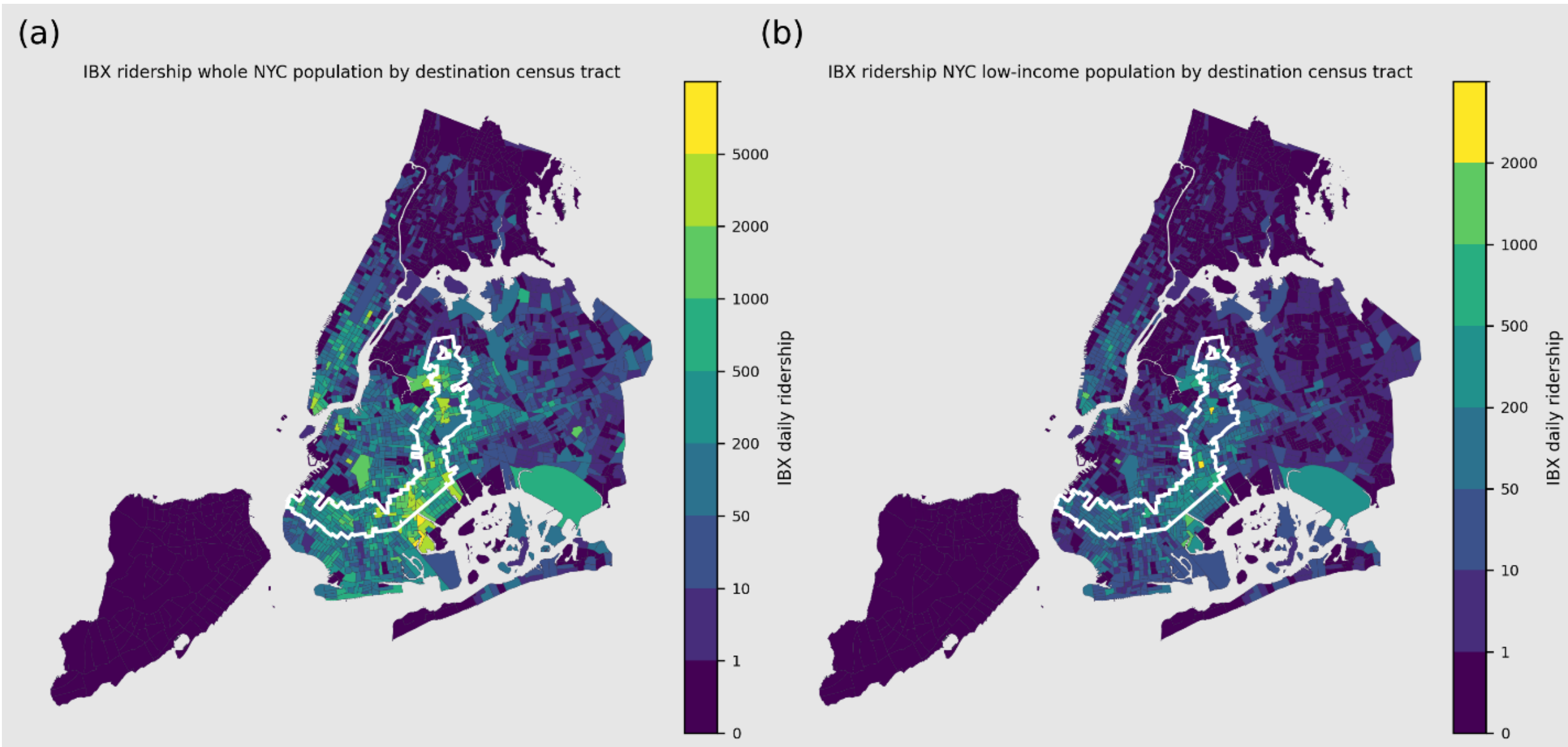

Figure 10. IBX daily ridership by destination census tract generated by (a) entire population and (b) low-income population.

Table 6. Summary of IBX impact on mode shift.

| Trip groups | Daily transit ridership increase (% total) | Daily Trips switch from private vehicle to transit (% total) | Potential GHG savings (metric ton/year) (% total) |
|---|---|---|---|
| **Whole NYC** | 40,277 (100%) | 16,449 (100%) | 30.63 (100%) |
| **Trip makers from low-income population** | 7,367 (18.30%) | 1,136 (6.91%) | 2.44 (7.97%) |
| **Departing from/arriving to the IBX corridor** | 27,798(69.01%) | 10,827 (65.82%) | 18.06 (58.97%) |

## 5.2. Consumer surplus benefit and equity analysis

The GLAM model facilitates quantification of the potential consumer surplus benefit attributed to the IBX, as determined by Eq. (10). The resulting consumer surplus benefits, expressed in monetary terms, are summarized in Table 7. For the whole population, the total consumer surplus benefit is estimated at 639.7 thousand US dollars daily which translates to $0.89 per trip. For trip makers from the low-income demographic, the total consumer surplus benefit is approximately 92.79 thousand US dollars, or $0.65 per trip made by low-income travelers. For travelers departing from or arriving at the IBX corridor areas, the average benefit gain is slightly higher at $0.7 per trip.

Table 7. Average consumer surplus benefit per trip brought by IBX.

| Trip group | Consumer surplus benefit (USD) per trip |
|---|---|
| **Whole NYC** | 0.89 |
| **Trip makers from low-income population** | 0.65 |
| **Departing from/arriving to the IBX corridor** | 0.70 |

To assess the equity impact of the IBX more comprehensively, the study employs both the CSDI and CSII. Regarding the level of disparity, the pre-IBX CSDI stands at 0.7588, signaling an inequitable distribution of mobility benefits between the low-income demographic and the population at large. With the IBX operational, the anticipated CSDI remains almost the same at 0.7582. This persistent disparity is caused by the significantly lower VOT among low-income populations (see Table 2). Because time-sensitive travelers value the IBX's time savings more highly, the inequity persists. Therefore, alongside improved transit convenience, targeted cost-reduction measures for low-income travelers are essential.

In the context of insufficiency analysis, the insufficiency threshold is initially defined as 10% of the average consumer surplus per trip. Prior to the IBX implementation, 33.56% of trips by low-income travelers yield a consumer surplus below this threshold. For the broader population, this figure is 21.29%, indicating a greater level of insufficiency among low-income travelers. Adjusting the threshold to 50% reveals that 63.54% of trips by low-income travelers result in a consumer surplus beneath the threshold, compared to 63.69% for all affected trips.

The introduction of IBX services slightly decreases insufficiency levels by less than 0.3%, signifying a negligible effect on equity from the standpoint of insufficiency. With a 10% insufficiency threshold, the pre-IBX CSII for the low-income group is 0.036, which slightly adjusts to 0.035 post-IBX introduction. The negligible difference between these values corroborates the findings from the insufficiency analysis. Increasing the threshold to 50%, the CSII shifts from 2.07 to 2.06 before and after the IBX's introduction, respectively. Overall, the IBX marginally improves the equity among low-income travelers from an insufficiency perspective.

The analysis demonstrates that the IBX does not fundamentally alter the equity landscape for low-income demographics. System-wide equity metrics, including CSII and CSDI, experience only marginal changes, indicating that the project is not a strong equity-driven measurement. Instead, higher-income groups continue to receive the most benefit. Furthermore, because only 15% of residents along the IBX corridor belong to a low-income group, these benefits are mathematically diluted by the larger population in equity evaluations. Based on this, we identify the IBX as equity enhancing but limited. The benefits are more concentrated toward localized residents, and the project can only address part of the larger transportation inequality issue.

## 6. Discussion and conclusion

In this study, we employ a straightforward methodology to predict and assess the potential impact of the newly proposed Interborough Express (IBX). This approach revolves around three core components: the NYC synthetic trip agenda, a set of NYC-specific choice model parameters, and anticipated changes in transit travel times for all OD pairs throughout the city. The NYC synthetic trip agenda maps out the daily activities and transportation mode choices of the entire NYC population on a typical workday, capturing current travel patterns across the city. The GLAM logit model, calibrated using the New York state synthetic population data from Replica, forecasts how travelers in each segment and OD pair might react differently to potential changes in transit travel times. We create a synthetic workday schedule for the IBX and use the Skim Matrices extension to estimate potential changes in transit travel times.

On average, individuals likely to use the IBX are projected to save about 28 minutes per transit trip, aligning with the savings proposed in the IBX project documents. Notably, trips originating or concluding near the IBX route exhibit greater time savings, underscoring the significant transit time reductions provided by this direct express link between communities in Brooklyn and Queens.

Using census tract OD-level transit time savings, the GLAM model estimates that over 272,000 trips would include the IBX in their routes, with more than 68% of these trips starting or ending near the IBX. This improved transit access and more direct routes would attract over 50,000 trips from other modes, with more than 40% of these shifted trips formerly made by driving private vehicles.

The predicted number of IBX-involved trips exceeds the 150,000 daily ridership estimate in the IBX proposal. However, as our study uses a relatively static method for estimating ridership without incorporating dynamic route choices, our projected numbers should be viewed as indicative of the potential scale of ridership rather than precise predictions. Nonetheless, our estimates generally align with the projections outlined in the IBX proposal.

A key advantage of utilizing a mode choice model lies in its ability to quantify potential consumer surplus benefits and analyze these benefits from various perspectives. The overall consumer surplus benefit from introducing IBX services is estimated at 0.89 USD per trip. Although the total benefit is substantial, it may not be uniformly distributed across different demographic groups. We employ two indices, the Consumer Surplus Disparity Index (CSDI) and the Consumer Surplus Insufficiency Index (CSII), to evaluate the equity impact of the IBX from a consumer surplus viewpoint. The CSDI, which assesses the disparity in consumer surplus between low-income groups and the broader population, shows marginally positive equity impacts brought by the IBX. The CSII is calculated to measure insufficiency levels. Like CSDI, the changes in CSII also suggests marginally positive equity impacts brought by the IBX.

This study evaluates potential short to medium term impacts based on current transportation networks and urban environments. As such, we assume that socio-demographic characteristics, travel patterns, and traveler evaluations of trip attributes remain static. This approach does not account for long term behavioral shifts driven by technological advancements, area level land use changes, or demographic shifts, nor does it consider infrastructure constraints such as rolling stock capacities.

Furthermore, by focusing on mode shift impacts based on aggregated trip performance, the analysis omits travel time dynamics caused by route choices and interchange impedances. While these microscopic behavioral shifts may be better understood in future studies using simulation techniques, our current findings offer a valuable baseline. Despite these limitations, the study provides significant insights into the scale of IBX's impact and serves as a meaningful reference for future transportation network analysis.

## Acknowledgments
This research was partially supported by the SEMPACT University Transportation Center USDOT #69A3552348302 and C2SMARTER Center USDOT #69A3552348326.

## Appendix

### A1. Validation of Replica's data

According to Replica's data quality report for the entire U.S. region[1], the largest error of demographic attributes for a single census tract unit is within 5% compared to census data, and the largest error of commute mode share for a single census tract unit is within 10% compared to CTPP data.

Tables A1 – A2 further compare inbound trip volumes and proportions to the Congestion Relief Zone (CRZ), using data from the New York Metropolitan Transportation Council (NYMTC)[2] and Replica. The year 2023 is selected because it is the year for which Replica's synthetic trip data are available. Table A1

[1] https://www.replicahq.com/data-validations. Retrieved January 21, 2026.
[2] https://www.nymtc.org/Data-and-Modeling/Transportation-Data-and-Statistics/Publications/Hub-Bound-Travel. Retrieved January 21, 2026.

examines inbound travel mode distributions, while Table A2 compares entry sector distributions across major corridors identified by NYMTC.

Overall, the two tables indicate a high degree of consistency between Replica and NYMTC data, with percent differences generally within 10% for most aggregate categories. The total inbound volume estimated by Replica is 6.54% lower than that reported by NYMTC, which is expected given that NYMTC data include truck trips whereas we excluded these trips from Replica data. Differences in non-motorized travel appear to arise primarily from mode classification, with some biking trips in NYMTC likely identified as walking in Replica; however, the combined volume of non-motorized trips remains comparable across the two datasets. These results suggest that Replica provides a reasonably accurate representation of inbound travel patterns to the CRZ in this study context, supporting its use for subsequent analysis of congestion pricing impacts.

**Table A1**
Comparison of inbound travel mode distribution: NYMTC vs. Replica Data

| | **NYMTC Data** | | **Replica Data** | | **Percent Difference** | |
|---|---|---|---|---|---|---|
| **Mode** | Volume (thousands) | Proportion (%) | Volume (thousands) | Proportion (%) | Volume (thousands) | Proportion (%) |
| Transit | 2057.87 | 70.96 | 1837.89 | 67.80 | -10.69 | -4.44 |
| *Subway* | 1494.26 | 51.54 | — | — | — | — |
| *Bus* | 215.78 | 7.42 | — | — | — | — |
| *Railroad* | 297.57 | 10.24 | — | — | — | — |
| *Ferry/Tram* | 51.26 | 1.76 | — | — | — | — |
| Auto (All) | 800.40 | 27.60 | 826.71 | 30.50 | +3.29 | +10.51 |
| *Driving* | — | — | 256.91 | 9.48 | — | — |
| *Carpool* | — | — | 358.72 | 13.23 | — | — |
| *On-demand* | — | — | 211.08 | 7.79 | — | — |
| Non-motorized | 42.01 | 1.45 | 46.11 | 1.70 | +9.77 | +17.44 |
| *Biking* | 42.01 | 1.45 | 21.77 | 0.80 | -48.18 | -44.56 |
| *Walking* | — | — | 24.34 | 0.90 | — | — |
| **Total** | **2900.28** | **100.00** | **2710.72** | **100.0** | **-6.54** | **+0.00** |

Note: Percent Difference = (Replica − NYMTC) / NYMTC × 100. NYMTC data is used as base. Indented rows show subcategories. "—" indicates data not available or not comparable. In NYMTC data, the category “Auto (All)” includes auto, taxi, van, and truck.

**Table A2**
Comparison of entry sector distribution: NYMTC vs. Replica data

| | **NYMTC Data** | | **Replica Data** | | **Percent Difference** | |
|---|---|---|---|---|---|---|
| **Entry Sector** | Volume (thousands) | Proportion (%) | Volume (thousands) | Proportion (%) | Volume (thousands) | Proportion (%) |
| 60th Street | 1071.24 | 36.94 | 982.47 | 36.24 | -8.29 | -1.87 |
| Brooklyn | 761.86 | 26.27 | 725.77 | 26.77 | -4.74 | +1.93 |
| Queens | 570.03 | 19.65 | 581.44 | 21.45 | +2.00 | +9.13 |
| New Jersey | 471.56 | 16.26 | 421.04 | 15.53 | -10.71 | -4.47 |
| Staten Island | 25.61 | 0.88 | — | — | — | — |
| **Total** | **2900.30** | **100.00** | **2710.72** | **100.00** | **-6.54** | **+0.00** |

Note: Percent Difference = (Replica − NYMTC) / NYMTC × 100. NYMTC data is used as base. "—" indicates data not available. The Staten Island Sector enters the CRZ via the Staten Island Ferry service and thus cannot be identified using Replica data.